\documentclass[prb,twocolumn,showpacs,preprintnumbers,amsmath,amssymb]{revtex4}
\usepackage{amsfonts}
\usepackage{graphicx}
\usepackage{dcolumn}
\usepackage{float}
\usepackage{bm}

\renewcommand{\theequation}{{\arabic{section}.\arabic{equation}}}
\makeatletter\@addtoreset{equation}{section} \makeatother

\newcommand{\ind}[1]{\textrm{#1}}
\newcommand{\com}[1]{}
\newcommand{\ket}[1]{{|#1\rangle}}

\newcommand{\user}[1]{{\cal #1}}

\begin{document}

\title{Unbalanced Renormalization of Tunneling
       \\in MOSFET-type Structures in Strong High-Frequency Electric Fields}

\author{Dmitry Solenov \footnote{E-mail: Solenov@clarkson.edu}}
\affiliation{Department of Physics, Clarkson University, Potsdam,
New York 13699--5820}


\begin{abstract}
Two-dimensional electron gas coupled to adjacent impurity sites in
high-frequency out-of-plane ac control electric field is
investigated. Modification of tunneling rates as a function of the
field amplitude is calculated. Nonlinear dependence on the ac
field strength is reported for the conductivity of two-dimensional
electron gas. It develops a periodic peak structure.
\end{abstract}


\maketitle

\section{Introduction}

Rapid development of experimental techniques at
nanoscale\cite{Jiang,Jiang2,Craig,Elzerman,Koppens} has stimulated
theoretical advances in describing quantum phenomena for various
geometries and settings. Extensive study has been done on
nonlinear effects in a few state quantum system subject to strong
harmonic
control,\cite{DunlapKenkre,GrossmannDittrich,GrossmannHanggi,DakhnovskiiBavli,
BavliMetiu,HolthausHone,RaghavanKenkre,HanggiReview,Burdov1,Burdov2,BurdovSolenov1,
BurdovSolenov2,BurdovSolenov3,Romanovs} such as a double quantum
dot,\cite{BurdovSolenov2} an array of coupled quantum
dots,\cite{BurdovSolenov3} superlatices,\cite{Romanovs} etc. In
this paper, we investigate the influence of the ac field on
one-state quantum objects coupled to two-dimensional electron gas
(2DEG) via tunneling.

The systems of such geometry have been recently used in
experimental as well as theoretical study of few and single
electron spin manipulations,\cite{Koppens} spin-to-charge
conversion measurements,\cite{Elzerman}
Ruderman-Kittel-Kasuya-Yosida (RKKY),\cite{Craig} and quantum
Hall\cite{Lewis} effects. In the mentioned experiments, 2DEG is
formed by confinement in a two-dimensional layer grown in a pile
between between the layers of the wide gap material. The electron
concentration in 2DEG can be varied significantly and is usually
controlled electrostatically by split-gate technique. A similar
system can be also created in the inversion layer in metal oxide
semiconductor field effect transistors (MOSFETs). In both cases,
the impurity centers (sites), localized usually outside the 2DEG
in adjacent layers, play an important role.\cite{Ralls,Mozyrsky}
One of the first experimental evidences here was the observation
of random telegraph noise in conduction of the inversion layer in
MOSFET.\cite{Ralls}

Surprisingly, little attention has been given to the control of
the impurity states, and thus the properties of 2DEG, dynamically.
Unlike the well-known phenomena of dynamical control of
tunneling\cite{HanggiReview,Burdov1,Burdov2,BurdovSolenov1,
BurdovSolenov2,BurdovSolenov3} in few state electron systems,
e.g., double quantum dot, the impurity-2DEG system provides more
degrees of freedom to change properties and correlations which are
not related to tunneling directly. One of the examples here is the
possible indirect influence over the RKKY interaction mediated by
2DEG electrons.\cite{Craig}

In what follows, we demonstrate that periodic high frequency
potential (electric field) applied perpendicular to 2DEG leads to
nontrivial renormalization and disbalance of the tunneling between
the impurity sites and 2DEG. Moreover, tunneling modification, as
well as Coulomb activation of the impurity sites, induces
oscillatory behavior of 2DEG conductivity as a function of the
amplitude of applied periodic field. This variation is similar to
Shubnikov--de Haas oscillations\cite{Engel} but have different
underlying physics.

In the next section, we formulate the impurity-2DEG model.
Section~\ref{Sec:FS} is devoted to the construction of the
corresponding stationary many-body problem using Floquet states.
Time-averaged quantities of interest are defined. In
Secs.~\ref{Sec:NST} and \ref{Sec:ST}, nonlinear dependence of
tunneling rates is obtained, starting with the simpler case that
neglects 2DEG electron scattering on impurity. The scattering
dynamics is analyzed. Finally, in Sec.~\ref{Sec:Cond}, field
amplitude dependence of conductivity is found. This dependence,
together with the expression of the tunneling rates, is the main
result of the paper.

\section{Model} \label{Sec:M}

\begin{figure}
\includegraphics[width=6.9cm]{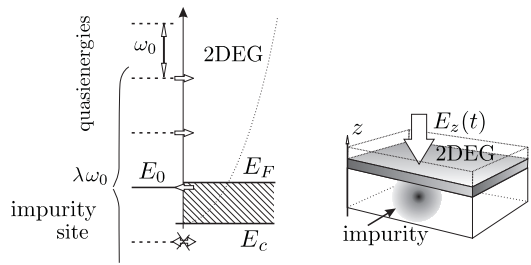}
\caption{Impurity site coupled to the 2DEG conduction electrons.
Strong ac field of frequency $\omega_0$ is applied perpendicular
to 2DEG. The ladder of quasienergies develops. All quasienergies
above $E_F$ and within the range of $\sim \lambda \omega_0$ are
used for tunneling \textit{out} process. The quasienergies below
the bottom of the conduction band, $E_c$, are not
active.}\label{fig1.eps}
\end{figure}

As mentioned above, we consider zero temperature 2DEG interacting
with an impurity electron localized in the adjacent layer of a
wide gap material. Both systems are subject to external plane
polarized harmonic field, $E_z(t)$, with the frequency $\omega_0$
and polarization along $z$ axis, perpendicular to the 2DEG plane,
see Fig.~\ref{fig1.eps}. The field is treated in the dipole
approximation appropriate as soon as the characteristic size of
the nanostructure is smaller compared to the field wavelength. The
impurity-2DEG interaction is via tunneling between 2DEG and
impurity states, as well as through Coulomb scattering of
conduction electrons on empty, positively charged, donor impurity
site. The electrons are considered spinless, and only the ground
state of the impurity site is taken into account. In many cases,
the heterostructure has more than one impurity site next to the
conduction layer. This situation, including the effect of spins,
is discussed in the last sections, and in most cases, the effect
of the external field can be deduced from the one-site spinless
model investigated below.

The unperturbed Hamiltonian is
\begin{equation}\label{Eq:M:H_0i}
H_{0,i}  = \frac{p_i^2}{2m} + U_d (\mathbf{r}_i) \delta_{i,d} +
U_\ind{2DEG}(z_i) \delta_{i,\mathrm{2DEG}},
\end{equation}
where $U_d(\mathbf{r})$ is the impurity localization potential,
and $U_\ind{2DEG}(z)$ denotes the potential profile that forms
2DEG. The $\delta$ functions are defined so that $H_{0,i}$
resemble the unperturbed Hamiltonian of an electron sitting on the
impurity ($\delta_{i,d}=1$, $\delta_{i,\mathrm{2DEG}}=0$) and in
the 2DEG ($\delta_{i,d}=0$, $\delta_{i,\mathrm{2DEG}}=1$). The
interaction between electrons as well as due to the external field
is
\begin{equation}\label{Eq:M:V}
V = - eE_z (t) \sum\limits_i {z_i}  + \sum\limits_{j\ne i}
\frac{e^2}{4\pi\epsilon_0\epsilon} \frac{ e^{ - q_s
|{\bf{r}}_i-{\bf{r}}_j|} }{|{\bf{r}}_i - {\bf{r}}_j|},
\end{equation}
where $E_z(t) = E_z \cos \omega_0 t$, and $q_s$ represents the
screening wave vector. Following the standard procedure, we define
the amplitudes
\begin{equation}\label{Eq:M:Delta}
H_{k0}^* = H_{0k}  = \int {d{\bf{r}}} \psi _0^* ({\bf{r}})
\left[H_{0,i}-\frac{p_i^2}{2m}\right] \psi _k ({\bf{r}})
\end{equation}
and
\begin{eqnarray}\label{Eq:M:Vkk}
V_{kk'}  &=&  - 2\int d\mathbf{r}d\mathbf{r}' |\psi
_0({\bf{r}})|^2
\\ \nonumber
&\times& \psi _k^* ({\bf{r}}')\psi _{k'}
({\bf{r}}')\frac{e^2}{{4\pi \epsilon _0 \epsilon }}\frac{{e^{ -
q_s \left| {{\bf{r}} - {\bf{r}}'} \right| } }}{{\left| {{\bf{r}} -
{\bf{r}}'} \right|}}.
\end{eqnarray}
Here, $\psi _0({\bf{r}})$ is the wave function of the
noninteracting ground state localized on the impurity, i.e.,
$H_{0,d}\psi _0({\bf{r}}) = E_0 \psi _0({\bf{r}})$, and $\psi
_k({\bf{r}})$ corresponds to the $k$-th state in 2DEG, i.e.,
$H_{0,\mathrm{2DEG}}\psi_k({\bf{r}}) = E_k \psi_k({\bf{r}})$. The
shape of $\psi_0({\bf{r}})$ and $\psi_k({\bf{r}})$ depends on the
form of $U_d({\bf{r}})$ and $U_\mathrm{2DEG}({\bf{r}})$. Though it
is not, strictly speaking, necessary, we will assume $H_{0k}$ to
be independent of $k$ and real, i.e., $H_{k0}^* = H_{0k} \to
\Delta$, to simplify notations. The electron-electron interaction
in 2DEG is ignored. With the above notations, we arrive to the
Hamiltonian
\begin{eqnarray}\nonumber
H &=& \left[ {E_0 + V_{00} (t)} \right]d^\dag  d + \sum\limits_k
{\left[ {E_k  + V_{2DEG} (t)} \right]} c_k^\dag  c_k
\\ \label{Eq:M:H}
&+& dd^\dag\sum\limits_{kk'} {V_{kk'} } c_k^\dag  c_{k'}  + \Delta
\sum\limits_k {\left( {d^\dag  c_k  + c_k^\dag  d} \right)},
\end{eqnarray}
where $V_{00}(t) = - e E_z(t) \int d\mathbf{r} \psi_0^*
({\bf{r}})z\psi _0({\bf{r}})$ and, similarly, $V_\ind{2DEG}(t)
\sim \int d\mathbf{r} \psi_k^* ({\bf{r}})z\psi _{k} ({\bf{r}})$.
The latter is assumed to be independent of $k$ since only the
distribution along $z$ of the lowest band is of interest. The $z$
distribution of $|\psi_0({\bf{r}})|^2$ and $|\psi_k({\bf{r}})|^2$
is concentrated around, respectively, the impurity center and the
middle of the $U_\mathrm{2DEG}$ quantum well. Therefore, $z_{0,0}
= \int d\mathbf{r} \psi_0^* ({\bf{r}})z\psi _0({\bf{r}})$ and
$z_{k,k} = \int d\mathbf{r} \psi_k^* ({\bf{r}})z\psi _{k}
({\bf{r}})$ refer to the position of the impurity and 2DEG along
the $z$ axis. As we will see later, only the difference between
these two quantities is of interest.

One can adopt the following values for order of magnitude
estimates. With the separation between the impurity and 2DEG of
the order of several angstroms, and the barrier height of the
order of several eV, the tunneling amplitude will vary by $\sim
1\div 10$ meV and smaller depending on the distance. The frequency
$\omega_0$ is of the order $\sim 10$ meV; the temperatures
$T\lesssim 1$ K; the size quantization $\gtrsim 100$~meV. These
values are feasible for Si/SiO$_2$ structures.

\section{Floquet States}\label{Sec:FS}

The Hamiltonian (\ref{Eq:M:H}) is periodic in time with the period
$2\pi/\omega_0$. It is natural to utilize this
symmetry.\cite{Shirley,Zeldovich,Ritus} Similar to space-periodic
solid state lattice structures, it was shown\cite{Zeldovich} that
the wave function corresponding to the periodic Hamiltonian is of
the form
\begin{equation}\label{Eq:FS:WaveFun}
\ket{\psi_\varepsilon(t)} = e^{-i\varepsilon t}\ket{u^\varepsilon
(t)},
\end{equation}
where $\ket{u^\varepsilon(t)}  = \ket{u^\varepsilon(t +
2\pi/\omega_0)}$, and $\varepsilon$ is the quasienergy. Moreover,
it was demonstrated that the set of quasienergy states can be
treated similarly to the conventional system of stationary
eigenstates---i.e., the system initially set up in a certain
quasienergy state (or distribution over these states) remains at
the same state (or with the same distribution) over the entire
evolution of the system.\cite{Zeldovich}

The transitions between the quasienergy states correspond to the
perturbations which break the periodicity. This has been
investigated in the literature.\cite{SolenovBurdov} In our case,
we assume the time scale of such perturbations due to environment
(or other factors) to be much larger than the one of interest. As
a result, one can analyze the quasienergy spectrum of the model to
obtain the information about tunneling effects in the system.

Let us show a few steps to support the above statement. It is
convenient to use the interaction representation, factoring out
the evolution due to the oscillatory part of the Hamiltonian
(\ref{Eq:M:H}). The corresponding evolution operator, $U_0 (t) =
\exp\{\frac{i}{{\omega _0 }}\sin \omega _0 t [ {V_{00} d^\dag d +
V_{2DEG} \sum_k {c_k^{^\dag  } c_k } } ]\}$, is still periodic so
that one can define
\begin{equation}\label{Eq:FS:MWaveFun}
\ket{\tilde \psi _\varepsilon  (t)}  = U_0^\dag  (t)\ket{\psi
_\varepsilon  (t)}  =  e^{i\varepsilon t} \ket{\tilde
u^\varepsilon (t)},
\end{equation}
with $\ket{\tilde u^\varepsilon(t)} = \ket{\tilde u^\varepsilon(t
+ 2\pi/\omega_0)}$. The corresponding Schrodinger equation is

\begin{equation}\label{Eq:FS:SchrEq}
i\frac{d}{{dt}}\ket{\tilde \psi _\varepsilon  (t)} = H(t)
\ket{\tilde \psi _\varepsilon  (t)},
\end{equation}
where
\begin{eqnarray}\nonumber
H(t) &=& E_0 d^\dag  d + \sum\limits_k E_k c_k^\dag  c_k +
dd^\dag\sum\limits_{kk'} V_{kk'} c_k^\dag  c_{k'}
\\ \label{Eq:FS:H_tilde(t)}
&+& \Delta \sum\limits_k {\left( { d^\dag c_k e^{-i\lambda\sin
\omega_0 t} + c_k^\dag d e^{i\lambda\sin \omega_0 t} } \right)}.
\end{eqnarray}
The effective strength of the external periodic field is defined
as $\lambda  = ( V_{00}  - V_\ind{2DEG} )/\omega_0$ with $V_{00}$
and $V_\ind{2DEG}$ representing the amplitudes of $V_{00}(t)$ and
$V_\ind{2DEG}(t)$ respectively. Note that $V_{00}$ and
$V_\ind{2DEG}$ have opposite signs and, thus, $\lambda$ is
proportional to the average distance $|z_{0,0}-z_{k,k}|$.

Our goal is to obtain the equation for quasienergy. One can easily
form equations for the Fourier transform of the periodic part of
the quasienergy wave function, $\ket{\tilde u^\varepsilon(t)}  =
\sum_m {e^{im\omega_0 t} } \ket{\tilde u^\varepsilon_m}$. From
Eq.~(\ref{Eq:FS:SchrEq}) we have
\begin{equation}\label{Eq:FS:QEEq}
\sum\limits_m {e^{ - im\omega_0 t} } \left[ {\varepsilon  + \omega
m - H(t)} \right]\ket{\tilde u^\varepsilon_{ - m}} = 0.
\end{equation}
Let us now recall that the time-dependent exponentials in
Hamiltonian (\ref{Eq:FS:H_tilde(t)}) have a simple series
representation $e^{i\lambda \sin \omega t}  =
\sum_{n=-\infty}^\infty{J_n (\lambda)e^{in\omega_0 t} }$ in terms
of the Bessel functions. It should be noted that the harmonic form
of the external field, and thus the time-dependent exponential in
Eq.~(\ref{Eq:FS:H_tilde(t)}), is not necessary. For any periodic
zero-average field, one can define the above series
representation. In this case, the Bessel functions $J_n(\lambda)$
are replaced with the coefficients, $J_n(\lambda) \to
f_n(\lambda)$, which carry the structure of a single oscillation.
The results obtained below will be qualitatively the same with
this modification. Since the harmonic field provides more insight
into the physics of the phenomenon, we use it in further
derivations instead of a more complicated time-periodic potential.

Defining a vector column of the states as $\ket{v} = \{
...,\ket{\tilde u_{ - 1}} ,\ket{\tilde u_0} ,\ket{\tilde u_1}
,...\} ^T$, we finally obtain the time-independent Schrodinger
equation for quasienergies,
\begin{equation}\label{Eq:FS:QESchrEq}
{\bf{H}}\ket{v}  = \varepsilon \ket{v},
\end{equation}
where the stationary Hamiltonian is
\begin{eqnarray}\nonumber
{\bf{ H}} &=& E_0 d^\dag  d + \sum\limits_k {E_k c_k^{^\dag } c_k
} + \omega_0 I_z  + dd^\dag\sum\limits_{kk'} V_{kk'} {c_k^\dag
c_{k'} }
\\  \label{Eq:FS:H_QE}
&+& \sum\limits_{n,k} {\Delta J_n (\lambda )\left( {I_ - ^{(n)}
c_k^\dag  d + I_ + ^{(n)} d^\dag  c_k } \right)}.
\end{eqnarray}
Here the additional operators are understood, if one defines a
column vector, $|e_m \rangle$, with all entries zeros except for
the $m$-th entry which is ``1." Then, $I_z |e_m \rangle  =
m|e_m\rangle$, $I_\pm |e_m \rangle  = |e_{m\pm 1}\rangle$. We use
the superscript in parentheses to generalize the power as $I_ \pm
^{( - n)} \mathop = \limits^{n
> 0} I_ \mp ^{(n)} \mathop = \limits^{n
> 0} I_ \mp ^n $. The quasienergy spectrum is now a solution to
the Kondo-type spin-assisted tunneling problem, where operators
$I_{\pm}$ correspond to renormalized rising (lowering) operators
of a large integer spin ($S \to \infty$), or an asymptotically
large ensemble of identical two-state systems. Note that
$I_{\pm}$, as they have been introduced to rewrite
Eq.~(\ref{Eq:FS:QEEq}) in the form (\ref{Eq:FS:QESchrEq}), are not
quite the spin rising (lowering) operators. Nevertheless, in the
limit of large spin ($S\to \infty$) and finite magnetization, they
differ only by a constant factor which has no effect on the
subsequent calculations.\cite{spincomment}

We should also demonstrate that the stationary problem with
Hamiltonian (\ref{Eq:FS:H_QE}) is sufficient to compute physically
observable quantities of interest. In this paper, we are after the
tunneling process in between the impurity and 2DEG, as well as the
conductivity of 2DEG, therefore, it is natural to investigate the
dynamics of the average occupation number for the impurity site,
i.e., $\overline{\langle \psi(t)|dd^\dag|\psi(t)\rangle}^t$, or
the amplitude of 2DEG electron transitions for states $k,k'$,
i.e., $\overline{\langle \psi(t)|c_k
c_{k'}^\dag|\psi(t)\rangle}^t$. The time-average is over the
period $2\pi/\omega_0$ of the fast external field oscillations.

Taking into account the properties of the quasienergy states, we
can focus on the average over a single quasienergy state. As
mentioned above, the average of two conjugate operators of the
same type is sought. This simplifies the expression further as
$\langle \psi_\epsilon(t)|dd^\dag|\psi_\epsilon(t)\rangle =\langle
\tilde\psi_\epsilon(t)|dd^\dag|\tilde\psi_\epsilon(t)\rangle$. The
time-averaged quantity becomes
\begin{equation}\label{Eq:FS:tAvdd}
\overline{\langle \tilde\psi_\epsilon(t)|
dd^\dag|\tilde\psi_\epsilon(t)\rangle}^{\,t} = \sum_m \langle
\tilde u_m |S(0,t) d(t)d^\dag(t) S(t,0)|\tilde u_{m} \rangle
\end{equation}
Here, the operators are in the interaction picture and evolve
according to the first three (main) terms of the Hamiltonian
(\ref{Eq:FS:H_QE}), while the standard scattering matrix is due to
the perturbation---the last two terms in Eq.~(\ref{Eq:FS:H_QE}).

As a result, the dynamics of the average occupation probability at
the impurity site is entirely determined by time-independent
Hamiltonian (\ref{Eq:FS:H_QE}). Similar arguments hold for the
transition amplitudes in 2DEG and thus the conductivity. A
standard equilibrium procedure of switching the interaction ``on"
adiabatically from $t=-\infty$ can be used. In this case the
initial dynamics is stationary in the first place, $|\tilde
\psi_\epsilon(t)\rangle=\mathrm{const}$, since Hamiltonian
(\ref{Eq:FS:H_tilde(t)}) becomes time-independent. This makes
$|\tilde u_{m} \rangle = |\tilde u_0\rangle \delta_{0,m}$. The
expression for the average becomes
\begin{eqnarray}\label{Eq:FS:MtAvdd}
&&\overline{\langle \tilde\psi_\epsilon(t)|
dd^\dag|\tilde\psi_\epsilon(t)\rangle}^{\,t} =
\\ \nonumber
&=& \langle \tilde u_0,e_0 |S(-\infty,t) d(t)d^\dag(t)
S(t,-\infty)|\tilde u_{0},e_0 \rangle,
\end{eqnarray}
where $|\tilde u_{0}\rangle$ is the usual initial state for
noninteracting fermions. In what follows, we will use the
shorthand notation for the complete average $\langle S(-\infty,t)
d(t)d^\dag(t) S(t,-\infty)\rangle$ instead of the one in
Eq.~(\ref{Eq:FS:MtAvdd}).

\section{Nonlinear tunneling without scattering}\label{Sec:NST}

In this section, we obtain the tunneling rates considering
Hamiltonian (\ref{Eq:FS:H_QE}) without Coulomb scattering on the
impurity, i.e., the fourth term. Let us explicitly show the main
part,
\begin{equation}\label{Eq:NST:H_0}
{\bf{ H}}_0 = E_0 d^\dag  d + \sum\limits_k {E_k c_k^{^\dag } c_k
} + \omega_0 I_z
\end{equation}
and the perturbation,
\begin{equation}\label{Eq:NST:V}
\mathbf{V} =\sum\limits_{n,k} {\Delta J_n (\lambda )\left( {I_ -
^{(n)} c_k^\dag  d + I_ + ^{(n)} d^\dag  c_k } \right)}.
\end{equation}

The perturbation (\ref{Eq:NST:V}) leads to equilibration of the
impurity occupation probability $P$. Using this state for
averaging, we have $\langle S(-\infty,\infty)\rangle = e^{i\phi}$.
As a result, the tunneling rates can be found by calculating the
zero-temperature impurity electron self-energy. The Green's
function $G(t,t')=-i\langle T d(t)d^\dag(t')
S(\infty,-\infty)\rangle$ is of interest, where
$S(t_2,t_1)=T\exp[-i\int_{t_1}^{t_2}dt'\mathbf{V}(t')]$.

For small $\Delta$, one-loop approximation is
sufficient.\cite{phonons} In Appendix~\ref{App:SepViaG}, we
calculate the self-energy for higher orders in $\Delta$. However,
they do not introduce any new physics and may be omitted. The
tunneling rate $\gamma$ is given by imaginary part of the
self-energy,
\begin{equation}\label{Eq:NST:SelfE}
\Sigma _1 (\omega ) = \sum\limits_{m,k} {\Delta ^2 } J_m^2
(\lambda )g_k (\omega  - m\omega _0 ),
\end{equation}
where $g_k(\omega)$ is noninteracting Green's function of 2DEG
electrons. The tunneling \textit{in} and \textit{out} from the
impurity can be clearly separated. The result is
\begin{equation}\label{Eq:NST:gamma_in_out}
\gamma_{in/out}= - 2\pi \Delta ^2 D_2 \sum\limits_{m =
M^1_{in/out}}^{M^2_{in/out}} {J_m^2 (\lambda )}.
\end{equation}
Here, $D_2$ is the density of states for 2DEG. The limits are
$M^1_{in}=-\theta(E_0-E_F)$, $M^2_{in}=-\sum_{n=1}^{\omega_0/E_F}
\theta(E_0-n\omega_0-E_F)$, and $M^1_{out}= \theta(E_F-E_0) +
M^2_{in}$, $M^2_{out}=\infty$, where $E_F$ is the Fermi energy of
2DEG. These limits and the summation terms have a clear physical
meaning. They correspond to tunneling \textit{in} and \textit{out}
from the impurity quasienergy states, see Fig.~\ref{fig1.eps}. At
the same time, they may be viewed in terms of allowed multiphoton
processes from the 2DEG state below the Fermi surface (tunneling
\textit{in}, negative $m$) and to the empty states above $E_F$
(tunneling \textit{out}, positive $m$). However, the latter
language should be used keeping in mind the explanation via the
quasienergies. The actual photons also account for the
renormalization of the tunneling amplitude, which is done
automatically in our treatment. The infinity in $M^2_{out}$ is
true in the limit sense, ${\lim}_{E_\infty \gg \omega _0 }
E_\infty/\lambda\omega_0$, and denotes the upper edge, $E_\infty$,
of the 2DEG band or the next conduction band if present relative
to the external field strength $\lambda \omega_0$.

To be specific, let us discuss the case when the electron
concentration in the conduction channel, $n_0$, is low, such that
the Fermi energy of 2DEG (with respect to the bottom of the
conduction band, $E_c$) is smaller as compared to the external
field frequency. The magnitude of the field exceeds the latter and
is much smaller than $E_\infty$. The chain inequality is $E_0 \sim
E_F < \omega _0 \lesssim \omega _0\lambda \ll E_\infty $, where
all the energies are measured from the bottom of the 2DEG
conduction band, $E_c$. This is a natural assumption for many 2DEG
systems used for few electron manipulation in recent experiments.
The tunneling rates are
\begin{equation}\label{Eq:NST:gamma_in}
\gamma _{in}  =  - 2\pi \Delta ^2 D_2 J_0^2 (\lambda )\theta (E_F
- E_0 )
\end{equation}
and
\begin{equation}\label{Eq:NST:gamma_out}
\gamma _{out}  = - 2\pi \Delta ^2 D_2 \sum\limits_{m = \theta
(E_F  - E_0 )}^\infty  {J_m^2 (\lambda )}.
\end{equation}
For a weak external harmonic field, $\lambda \ll 1$, both
tunneling rates approach the well known result $\gamma_0  =  -
2\pi \Delta ^2 D_2\theta( \pm E_F \mp E_0 )$. This is also true
for the more general case given in
Eq.~(\ref{Eq:NST:gamma_in_out}). The low amplitude of the field
suppresses the multiphoton absorption (emission) process,
$J_{m>1}^2(\lambda)\to 0$, and allows the single photon processes
as a small first-order correction, $J_{1}^2(\lambda)$, while the
renormalization of the non-assisted tunneling vanishes,
$J_0^2(\lambda) \to 1$.

Larger external field amplitudes activate more quasienergy states.
Keeping in mind the earlier discussion, this process can also be
viewed in terms of induction of multiphoton transitions with the
maximal number of photons absorbed (emitted) per transition $\sim
\lambda$. In this case, the tunneling rates depart from each
other, see Fig.~\ref{fig2.eps}. In the limiting case of
$\lambda\gg 1$, the tunneling \textit{out} rate becomes
$\gamma_{out}  \to  - 2\pi \Delta ^2 D_2 \frac{1}{2}$, while
$\gamma_{in}$ oscillates according to $J^2_0(\lambda)$ and
converges to zero as $\sim 1/\lambda$. Similar results can be
obtained directly by averaging the transition matrix element due
to Hamiltonian (\ref{Eq:M:H}) over the fast oscillations of
driving field, keeping the terms of the order $\Delta^2$, see
Appendix~\ref{App:AveExSol}. The above approach, however, provides
more physical insight and is more convenient for further
discussion.

It should be noted that the above results apply in a more general
case when $E_0$ is significantly higher or lower then $E_F$. In
this case one can define $\bar{E}_0 = E_0 + \bar{m}\omega_0$, such
that $|\bar{E}_0 - E_F|<\omega_0$. Different quasienergy comes
near the resonance with the Fermi surface, see
Fig.~\ref{fig1.eps}. The same form of the expression for the
tunneling rates, Eqs.~(\ref{Eq:NST:gamma_in}) and
(\ref{Eq:NST:gamma_out}), can be used by replacing $E_0 \to
\bar{E}_0$ and $J_m(\lambda) \to J_{m + \bar{m}}(\lambda)$.

\section{Scattering Dynamics} \label{Sec:ST}

Let us now investigate the modification of tunneling rates due to
scattering, i.e., the fourth term, $V_{sc}$, of the Hamiltonian
(\ref{Eq:FS:H_QE}), ignored in the previous section. This problem
is similar to the problem of x-ray edge
singularity.\cite{NozieresDominics} To produce a tractable
solution, one has to assume a special form of the scattering
potential, $V_{kk'} \to V u_k u_{k'}$. Here, $u_{k}$ is the cutoff
function which is of the order ${\cal O}(1)$ for $E_{k} \sim E_F,
\lambda \omega_0$ and vanishes for $E_{k} \sim E_\infty$. We also
assume $u_k$ to be symmetric.
\begin{figure}
\includegraphics[width=6.5cm]{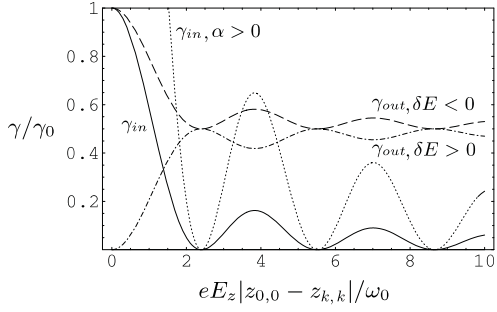}
\caption{Tunneling rates as a function of harmonic field
amplitude. The rates are given in terms of zero-field tunneling
rate amplitude $\gamma_0$. The dotted curve represents the
tunneling \textit{in} rate with singular renormalization factor
due to scattering with $\alpha>0$. In this case, only qualitative
dependence is presented, since the scattering factor will involve
regularization, as discussed in the text.}\label{fig2.eps}
\end{figure}

One can notice that the scattering is only present when the
electron leaves the impurity site.\cite{NozieresDominics} This
differentiates the tunneling \textit{in} and \textit{out}
processes. To investigate the two in a uniform treatment, it is
convenient\cite{NozieresDominics} to redefine the perturbation due
to scattering for tunneling \textit{in} as $V_{sc} = - V d^\dag d
\sum_{kk'}u_k u_{k'}c_k^\dag c_{k'}$. We add the corresponding
term to the unperturbed Hamiltonian, i.e., $\mathbf{H}_0 \to
\mathbf{H}_0 + \sum_{kk'}u_k u_{k'} c_k^\dag c_{k'}$. This will
only redefine the noninteracting conduction electron Green's
function, $\tilde{g}_{kk'}(\omega) = g_{k}(\omega)\delta_{kk'} - i
V \sum_{k''} g_{k}(\omega) u_k u_{k''} \tilde{g}_{k''k'}(\omega)$.
For tunneling \textit{out}, one still has $V_{sc} = V dd^\dag
\sum_{kk'}u_k u_{k'}c_k^\dag c_{k'}$.

To utilize the results of Ref.~\onlinecite{NozieresDominics}, note
that the tunneling rates can also be obtained by calculating the
time derivative of $\langle T d(t)d(t \pm
0)S(\infty,-\infty)\rangle$. The expression correct up to
$\Delta^2$ is
\begin{eqnarray}\label{Eq:ST:I_tmp}
&&\gamma _{in/out}  = {\mathop{\rm Re}\nolimits} \sum\limits_{n,k}
{\Delta ^2 J_n^2 (\lambda )}
\\ \nonumber
&\times& \int\limits_{ - \infty }^\infty {dt'} \langle |Tc_k^\dag
(t)d(t)d^\dag  (t')c_k (t')|\rangle \langle I_ - ^{(n)} (t)I_ +
^{(n)} (t')\rangle.
\end{eqnarray}
Here, the state $|\rangle$ include the $|0\rangle$ or $|1\rangle$
state of the impurity for the tunneling \textit{in} or
\textit{out} cases, respectively. The problem is to compute the
average
\begin{equation}\label{Eq:ST:F(t)}
F(t - t') = \sum_{kk'} {u_k u_{k'} } \langle T\bar{c}_k^\dag
(t)\bar{d}(t)\bar{d}^\dag  (t')\bar{c}_{k'} (t')\rangle.
\end{equation}
Then, the rates are found from $F(t)$ via its Fourier transform as
\begin{equation}\label{Eq:ST:I}
\gamma _{in/out}  = {\mathop{\rm Re}\nolimits} \sum\limits_{n,k}
{\Delta ^2 J_n^2 (\lambda )} F(n\omega _0 ).
\end{equation}
With the above redefinition of the scattering perturbation, the
average (\ref{Eq:ST:F(t)}) can be computed for any magnitude of
the scattering amplitude as a one-body problem with time-dependent
potential (scattering on impurity), as it was demonstrated by
Nozieres and De Dominics.\cite{NozieresDominics} This is possible
since we assume that the impurity has no internal degrees of
freedom. If one defines the times of two tunneling acts by $t$ and
$t'$, the average (\ref{Eq:ST:F(t)}) is found via the time
evolution of $\varphi_{kk'}(\tau,\tau',t,t')=\langle T
\bar{c}_k(\tau) \bar{c}_{k'}^\dag(\tau') \rangle$ with all the
vertices describing the scattering acts restrained to the interval
$(t,t')$. The overbar denotes complete evolution.

When the region around the Fermi energy is of interest, $|\delta
E| \ll \omega_0$ and $m=0$, the asymptotic form of the scattering
can be used.\cite{NozieresDominics} This adds a singular factor to
$F(\omega \to 0)$, and thus the tunneling rate, of the form
\begin{equation}\label{Eq:ST:SingFactor}
\left[{\xi_0}/{(\pm \delta E)}\right]^\alpha,
\end{equation}
where $\alpha = 2\delta/\pi - (\delta/\pi)^2$ and $\delta E =
E_F-E_0$. In two dimensions the phase shift is defined by $\tan
\delta = \pi D_2 V$. Here, $\xi_0$ is the cutoff coming from
$u_k$. The exponent in Eq.~(\ref{Eq:ST:SingFactor}) is found
within logarithmic accuracy.

In our case, higher energy terms are present. They do not comply
with the asymptotic approximation for the scattered wave functions
used to obtain Eq.~(\ref{Eq:ST:SingFactor}). For higher energies,
$m\neq 0$, a short-time dynamics of $F(t)$ is necessary. In the
case when $Vt\ll 1$, it is possible to omit the integral term in
Eq.~(35) of Ref.~\onlinecite{NozieresDominics}. In other words,
the scattering becomes less important. As the result, the
corresponding tunneling terms are the same as in the previous
section up to ${\cal O}(V/m\omega_0)$ corrections which are
negligible provided $V/\omega_0\ll 1$. The tunneling due to large
$V$ is also clear. The tail of the scattering renormalization will
be added, as a factor, until approximately the $n \sim V/\omega_0$
tunneling term.

Finally, for the case of Eqs.~(\ref{Eq:NST:gamma_in}) and
(\ref{Eq:NST:gamma_out}) and assuming that $|\delta E| \ll
\omega_0$, we obtain
\begin{equation}\label{Eq:ST:gamma_in}
\gamma _{in}  =  - 2\pi \Delta ^2 D_2 J_0^2 (\lambda )\theta
(\delta E)\left( {\frac{{\xi _0 }}{{\delta E }}} \right)^\alpha
\end{equation}
and
\begin{eqnarray}\label{Eq:ST:gamma_out}
\gamma_{out}  = &-& 2\pi \Delta ^2 D_2 J_0^2(\lambda )
\theta(-\delta E)\left(\frac{\xi _0}{-\delta E}\right)^\alpha
\\ \nonumber
&-& 2\pi \Delta ^2 D_2 \sum_{m = 1}^\infty J_m^2(\lambda ).
\end{eqnarray}
When the external harmonic field is weak the tunneling rates again
approach the standard expression (with the scattering
renormalization). For strong fields, the result depends on the
scattering exponent as well as on $\delta E$, unlike in
Eqs.~(\ref{Eq:NST:gamma_in}) and (\ref{Eq:NST:gamma_out}). The
above solution is not valid for intermediate values of $\delta E$,
nevertheless its possible form is rather clear and can be inferred
from Eqs.~(\ref{Eq:ST:gamma_in},~\ref{Eq:ST:gamma_out}) as far as
the external field influence is concerned.

As it was mentioned in the previous section, the impurity energy
level may be well below (or above) the 2DEG Fermi energy. In this
case, a different quasienergy enters the resonance with $E_F$
resulting in the singular renormalization of the corresponding
term. The obtained result applies with the replacements $E_0 \to
\bar{E}_0$ and $J_m(\lambda) \to J_{m + \bar{m}}(\lambda)$, where
$|\bar{E}_0 - E_F|\ll \omega_0$.

The singular factor leads to the two effects, similar to those
discussed in Ref.~\onlinecite{NozieresDominics} for x-ray
absorption (emission) problem. It destroys the jump in energy
dependence for tunneling rate if $\alpha <0$. When $\alpha >0$,
the jump becomes larger but is finite due to the presence of the
spin degrees of freedom and temperature broadening of the electron
density near the Fermi energy which tend to quench the
singularity. In the $\alpha <0$ case, the difference between
\textit{in} and \textit{out} tunneling rates is stronger; the
higher tunneling \textit{in} rate (see Fig.~\ref{fig2.eps}) makes
$\gamma_{in/out}$ rates closer near the resonance, except for the
values of $\lambda$, such that $J_0(\lambda)=0$, in which case the
singularity is suppressed. Due to this latter fact, one obtains
sharp peaks in the low-temperature resistivity of 2DEG as a
function of $\lambda$, as will be shown in the next section. For
both cases of $\alpha$, the difference between $\gamma_{in/out}$
vanishes for high temperatures, since multiphoton transitions to
the high energy region become possible in both ways.

\section{Conductivity of 2DEG}\label{Sec:Cond}

Let us analyze how the strong-field modification of tunneling
affects the conductivity of 2DEG. Since the oscillating electric
field is perpendicular to the 2DEG plane, it should not influence
the conduction electrons directly, but only through the
interaction with the adjacent impurities. We assume that the
impurities are distributed with the density $n_i$, small enough to
neglect the interference between scattering on different
sites,\cite{equilibration} as well as the tunneling between the
impurity sites. The contribution due to other scattering processes
are not of interest here. They are not affected by the field in
our system and will add $\lambda$-independent terms to the total
resistivity. The conductivity is calculated as a linear response
to a vanishingly small in-plane dc electric field.

Conductivity due to impurity scattering is given by $\sigma  = e^2
n_0 \tau(E_F)/m$, where $e$, $n_0$, and $m$ are the elementary
charge, electron concentration, and effective mass, respectively.
The scattering time $\tau(E_F)$ can be estimated with the Green's
function relaxation time. The difference between the two is a well
known $(1-\cos\theta ')$ factor.\cite{Mahan} As far as the field
influence is concerned, one can estimate $\tau(E_F)$ by evaluating
the imaginary part of the retarded self-energy of conduction
electrons,\cite{scattimecomment} i.e., $\tau^{-1}(E_F) \sim -
\mathrm{Im}\Sigma_\mathrm{ret}(k_F ,\omega \to 0)$.

For a dilute impurity system,\cite{Mahan}
\begin{eqnarray}\label{Eq:Cond:S-E_exp}
\Sigma _k^{\rm{sc}}(i\omega_n)\!\! &=& n_i \left\{ V_{kk}  +
\frac{1}{\nu} \sum_{k'} V_{kk'} V_{k'k} \user{G}_{k'} (i\omega _n
) \right.
\\ \nonumber
&+&\!\!\!\! \left. \frac{1}{{\nu ^2 }}\!\sum\limits_{k'k''}
{V_{kk'} V_{k'k''} V_{k''k}
\user{G}_{k'} (i\omega _n )\user{G}_{k''} (i\omega _n )}  + ...
\right\},
\end{eqnarray}
where $\user{G}_{k} (i\omega _n )$ is the Matsubara Green's
function of 2DEG electrons. In the limit of small concentrations,
$n_i \to 0$, the noninteracting function [without $\Sigma
_k^{\rm{sc}}(i\omega_n)$] may be used. Assuming $V_{kk'} \to Vu_k
u_{k'}$ as before, for 2D electron system with stationary impurity
scatterers, one has\cite{scattimecomment2} $\mathrm{Im}\Sigma
_{\rm{ret}}^{\rm{sc}} (k_F,\omega \to 0) = - \frac{{n_i }}{{\pi
D_2 }}\sin ^2 \delta$.

The tunneling affects the equilibrium occupation of the impurity,
as well as $\user{G}_{k} (i\omega _n )$. The scattering vertex is
modified as $V_{kk}  \to  - V_{kk} \langle dd^\dag \rangle = -
V_{kk} [1 - P(\lambda)]$ for the donor impurity site [for the
acceptor site, one has $V_{kk}  \to  - V_{kk} P(\lambda)$]. We
note that for the equilibrium state, the averages are over $\ket{}
= \sqrt P(\lambda) \ket{}_1 + \sqrt {1 - P(\lambda)} \ket{}_0$,
where $\ket{}_1$ corresponds to the filled impurity site and
$\ket{}_0$ to the empty site. The occupation probability is
\begin{equation}\label{Eq:Cond:P}
P(\lambda) = \frac{\gamma_{in}(\lambda)}{\gamma_{out}(\lambda) +
\gamma_{in}(\lambda)}.
\end{equation}

The 2DEG electron Green's function becomes $\user{G}_k(i\omega _n
) = [\user{G}_{0,k}(i\omega _n )^{- 1} - \Sigma_k^t(i\omega
_n)]^{-1}$, where $\user{G}_{0,k}(i\omega _n)$ and
$\Sigma_k^t(i\omega _n)$ are noninteracting Green's function of
conduction electrons and the corresponding self-energy due to the
tunneling potential (\ref{Eq:NST:V}), respectively. The latter is
$\Sigma _k^t (i\omega _n ) = \sum_m {\Delta ^2 } J_m^2 (\lambda
)\user{G}(i\omega _n  - m\omega _0 )$. For the impurity scattering
self-energy, it is sufficient to use the noninteracting function
$\user{G} \to \user{G}_0$ in $\Sigma_k^t(i\omega_n)$.

When calculating the tunneling contribution to conductivity, the
effect of scattering in $\Sigma_k^t(i\omega_n)$ is included as
suggested earlier. This results in the factor of $(\delta
E/\xi_0)^{(\delta/\pi)^2}/\delta E$ for the term with $m=0$, with
the divergence, at $\delta E \to 0$ suppressed, as discussed
above. The renormalization of $\Delta$ at resonance, i.e., $\Delta
\to \kappa \Delta$, takes place. The renormalized term, however,
may be suppressed by the choice of $\lambda$, such that
$J_0(\lambda)=0$.
\begin{figure}
\includegraphics[width=7.0cm]{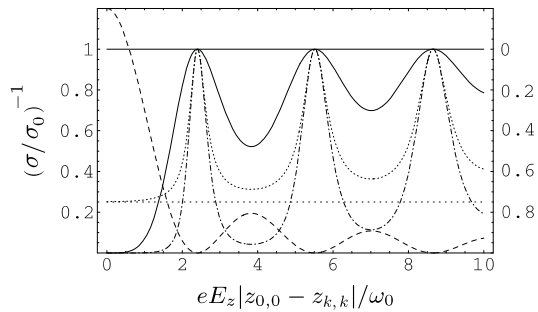}
\caption{Reciprocal conductivity as a function of harmonic field
strength. The right complementary scale is for acceptor impurity
sites. The solid curve is $\sigma^{-1}_{sc}$ for $\delta E>0$. The
horizontal solid line represents $\sigma^{-1}_{sc}$ for $\delta
E<0$, when the impurity sites are empty. The dashed curve shows
$\sigma^{-1}_{t}$ up to a factor of $\kappa^2$. In this case, only
the left scale is appropriate. The dash-dotted curve represents
$\sigma^{-1}_{sc}$ for $\delta E>0$ with the singular
renormalization factor with $\alpha>0$. For stronger
amplifications, the curve follows closer to zero in its minimums.
The dotted curve gives a qualitative dependence for finite
temperatures at $\delta E \to 0$, $\alpha>0$. In this case, the
reciprocal conductivity curve becomes bounded from the bottom
(dotted horizontal line)}\label{fig3.eps}
\end{figure}
\begin{figure}
\includegraphics[width=7.0cm]{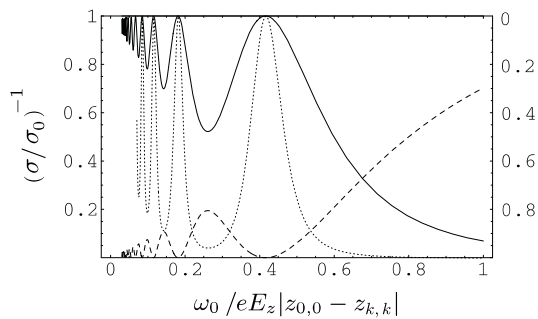}
\caption{Reciprocal conductivity as a function of harmonic field
frequency, $\omega_0$. The right complementary scale is for
acceptor impurity sites. The curves are cut at low frequency to
exclude heavy oscillations. The solid curve is $\sigma^{-1}_{sc}$
for $\delta E>0$. The dashed curve is $\sigma^{-1}_{t}$ up to a
factor of $\kappa^2$ (only the left scale is appropriate in this
case); the dotted curve corresponds to the dash-dotted line in
Fig.~\ref{fig3.eps}.}\label{fig4.eps}
\end{figure}

Finally, we have two contributions to the conductivity. One is due
to the tunneling,
\begin{equation}\label{Eq:Cond:CondT}
\sigma_{t}^{-1} \sim \tau_t(E_F)^{-1} \sim \Delta^2 J_0^2(\lambda)
\delta'(\delta E),
\end{equation}
where the $\delta$-function $\delta'(x)$ is broadened so that
$\delta'(0)=\kappa^2$. The other is caused by scattering,
\begin{eqnarray}\label{Eq:Cond:CondSC}
\sigma_{sc}^{-1} &\sim& \tau_{sc}(E_F)^{-1}
\\ \nonumber
&\sim& \frac{n_i \tan^2(\delta) \gamma_{in}^{-2}(\lambda)
\gamma_{out}^2(\lambda)}
{\gamma_{in}^{-2}(\lambda)\gamma_{out}^2(\lambda)(\cos\delta)^{-2}
+ \frac{2 \gamma_{out}(\lambda)}{\gamma_{in}(\lambda)}+1}.
\end{eqnarray}

In Figs.~\ref{fig3.eps} and \ref{fig4.eps}, we plot the reciprocal
conductivity due to tunneling resonance,
Eq.~(\ref{Eq:Cond:CondT}), and scattering,
Eq.~(\ref{Eq:Cond:CondSC}), as a function of $\lambda$ and
$1/\lambda$. The result is given in terms of the conductivity
$\sigma_0$ due to the scattering on stationary ionized impurities
in the absence of external harmonic field. The tunneling
contribution to resistivity, $\sigma_{t}^{-1}$, features the two
state aspect of the impurity-2DEG coupling. It reflects the
dynamical suppression of resonant tunneling (dynamical
localization) similar to the double quantum dot
systems\cite{BurdovSolenov1,BurdovSolenov2,BurdovSolenov3,SolenovBurdov}
where the tunneling is suppressed by $J_0(\lambda)$ (or
$J_{\bar{m}}(\lambda)$ for lower $E_0$ or biased structures) as
well. This contribution is proportional to $\Delta^2 \kappa^2$ and
vanishes for large fields as $1/\lambda$. It is independent of the
donor (acceptor) type of the impurity.

The solid curve in Fig.~\ref{fig3.eps} and \ref{fig4.eps} show the
resistivity due to scattering of conduction electrons on
tunneling-active impurities, $\sigma_{sc}^{-1}$. The modification
of tunneling is not considered in this case, and the corresponding
rates are given by Eqs.~(\ref{Eq:NST:gamma_in}) and
(\ref{Eq:NST:gamma_out}). When $\delta E<0$, the impurity sites
are empty at equilibrium and the scattering occurs with the
highest probability---the conductivity is not affected by the
field. When $\delta E>0$, $\sigma_{sc}^{-1}$ oscillates as a
function of $\lambda$. For $\lambda=0$, the occupied impurities do
not scatter 2DEG electrons (the situation is opposite for acceptor
sites). At $\lambda \gg 1$ tunneling \textit{out} transitions from
higher quasienergies dominate, leaving the impurity empty. The
values of resistivity corresponding to $\delta E>0$ and $\delta
E<0$ converge to each other, see Fig.~\ref{fig4.eps}. In the
figure, $\tan\delta = 0.1$. For larger values of $V$ the
oscillations converge to ``1" faster.

When the singular renormalization is introduced, the resistivity
peaks become sharper for $\alpha>0$ and $\delta E>0$. The
singularity amplifies the tunneling to the impurity site for all
$\lambda$ except for near the zeros of $J_0(\lambda)$,
deactivating the scatterers. The corresponding curves in
Figs.~\ref{fig3.eps} and \ref{fig4.eps} are given for
amplification factor of 10. The estimate for small (but finite)
temperature with $\delta E \to 0$ is shown as a dotted curve in
Fig.~\ref{fig3.eps}. In this case, the horizontal line gives the
lower bound for large amplifications.

We have investigated tunneling and conductivity modification in
impurity-2DEG structure in external time-periodic field. Nonlinear
dependence on the field amplitude has been obtained and analyzed
for both tunneling rates and conductivity. The calculations have
been performed in the limit of small $\Delta$ and nearly zero
temperatures. Further investigation has to be done to understand
the modification of other 2DEG electron correlations, such as the
ones leading to RKKY coupling. The presence of larger currents is
also of interest.


\acknowledgments

The author acknowledges stimulating discussions with G. F. Efremov
and V. Privman, and funding by the NSF under Grant No.
DMR-0121146.

\appendix
\renewcommand{\theequation}{\thesection .\arabic{equation}}

\section{}\label{App:SepViaG}

We need to analyze the decay of $\langle S(-\infty,t)
d(t)d^\dag(t) S(t,-\infty)\rangle$. The scattering matrix is due
to the interaction (\ref{Eq:NST:V}), with the unperturbed part
(\ref{Eq:NST:H_0}). Define the equilibrium Green's function
$G(t,t')=-i\langle T d(t)d^\dag(t') S(-\infty,\infty)\rangle$.
Then, the desired average is $G(t,t'\to t \pm 0)$.

Since $\mathbf{H}_0$ and $\mathbf{V}$ are time-independent, one
can obtain the dynamics by differentiating $G(t,t')$ with respect
to one of the times. Inverting the operator $i\partial/\partial t'
- E_0$, one obtains
\begin{equation}\label{Eq:AppG:Geq}
G(t-t'')\! = g(t-t'') - i \Delta \!\!\ \sum_{m,k}
\int\limits_{-\infty }^{\infty}\!\!\! dt' M_m(t,t')g(t'-t''),
\end{equation}
where
\begin{eqnarray}\label{Eq:AppG:M}
M_m(t,t') = J_m (\lambda )\sum\limits_{n} {( - i)^n }
\int\limits_{ - \infty }^\infty  {dt_1 ...dt_n }
\\ \nonumber
\times \langle T d(t)V(t_1 )...V(t_n )c_{k'}^\dag  (t')I_ - ^{(m)}
(t')\rangle.
\end{eqnarray}
Here, summation over the connected diagrams is assumed. It is
easily noticed that only odd-$n$ terms will contribute to the
expression. The first-order term contains the average $\langle
Td(t)I_ + ^{(n)} (t_1 )d^\dag  (t_1 )c_k (t_1 )c_{k'}^\dag  (t')I_
- ^{(m)} (t')\rangle$ which splits as
$g(t-t_1)g_k(t_1-t')g_m(t_1-t')$, where $g(t-t_1)$ and
$g_k(t_1-t')$ are noninteracting Green's functions for impurity
and 2DEG electrons, respectively, while $g_m(t_1-t') =
e^{im\omega_0(t-t')}$. The corresponding diagram is
\begin{figure}[H]
\begin{center}
\includegraphics[width=1.5cm]{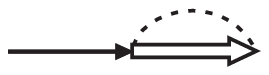}
\end{center}
\end{figure}
{\noindent}where the solid arrow is $g(t-t_1)$, the double arrow
is $g_k(t_1-t')$, and the dash-arrow represents $g_m(t_1-t')$. The
next-order, $n=3$, terms have two $g(t_1-t_2)$ and $g_k(t_1-t_2)$
with four spin vertices $I_\pm^{(n)}$. Some of the diagrams are
\begin{figure}[H]
\begin{center}
\includegraphics[width=7.0cm]{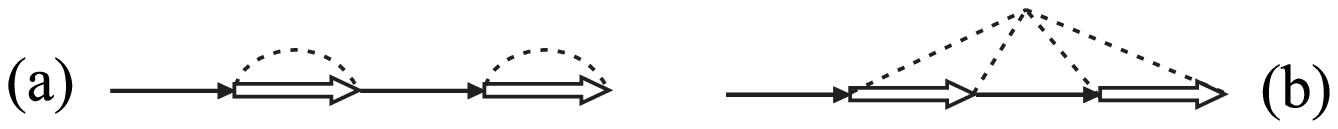}
\end{center}
\end{figure}
{\noindent}The second (b) diagram represents the fact that
averages $\langle I_ + ^{(n_1 )} (t_1 )I_ - ^{(n_2 )} (t_2 )I_ +
^{(n_3 )} (t_3 )I_ - ^{(m)} (t')\rangle$ do not necessarily split
in pairs---the only restriction is $n_1-n_2+n_3-m=0$. The diagrams
of type (a) contribute to a single loop approximate solution,
\begin{equation}\label{Eq:AppG:G_1loop}
G(\omega ) = g(\omega ) + g(\omega )\Sigma _1 (\omega )G(\omega ),
\end{equation}
where the self-energy is
\begin{equation}\label{Eq:AppG:SelfEnergy}
\Sigma _1 (\omega ) = \sum\limits_{m,k} {\Delta ^2 } J_m^2
(\lambda )g_k (\omega  - m\omega _0 ).
\end{equation}
This result corresponds to the first-order self-energy of the
general solution. The other terms can also be evaluated.

Note that operators $I_{\pm}$ describe an asymptotically large
spin\cite{spincomment} and thus commute with each other when
averaged over the states corresponding to zero (finite)
magnetization, see Eq~(\ref{Eq:FS:MtAvdd}). Therefore, one can
replace $I_{\pm}^{(n)}(t) \to e^{\pm in\omega_0 t}$ with the
restriction that the sum of all $\pm n$ in the expression equals
zero. After some algebra, the $N$-th order ($N>1$) self-energy can
be obtained in the form
\begin{eqnarray}\nonumber
\Sigma _N (\omega ) &=& \!\!\!\!\! \sum\limits_{n_1,...,n_{2N}
,k_1,...,k_N } \!\!\!\! {\Delta ^{2N} } J_{n_1 } (\lambda )J_{n_2
} ( - \lambda )...
\\ \label{Eq:AppG:SelfEnergyN}
&\times& \delta (n_1  + n_2  + ... + n_{2N - 1}  + n_{2N} )
\\ \nonumber
&\times& g_{k_1 ,n_2 ,n_{3,4} ,...} (\omega )g_{n_{3,4} ,...,n_{2N
- 1,2N} }(\omega )...
\\ \nonumber
&\times& g_{n_{2N - 1,2N} } (\omega )g_{k_N ,n_{2N} } (\omega ),
\end{eqnarray}
where the sign of summation indices entering the expression as
$-n$ has been changed, $n_{ij} = n_i + n_j$. We have defined
$g_{\xi ,\xi',...}(\omega ) = g(\omega  - \omega _0 \xi  - \omega
_0 \xi ' - ...)$ and $g_{k,\xi ,\xi ',...} (\omega ) =
g_{k}(\omega - \omega _0 \xi  - \omega _0 \xi ' - ...)$. The
summation in Eq.~(\ref{Eq:AppG:SelfEnergyN}) should not include
the terms which have $g_{(k),\xi ,\xi',...}(\omega )$ with
$\xi+\xi'+...=0$. These terms are taken into account by the
preceding self-energies. An estimate of the above sum suggests
that $\Sigma _N \sim x^N$, with $x=\Delta^2 D_2/\omega_0$. In our
case, $\Delta^2 D_2/\omega_0 \sim \Delta^2/\omega_0 E_F \ll 1$.
Therefore, a one-loop approximation is sufficient.

\

\

\section{}\label{App:AveExSol}

Let us obtain the first-order tunneling rate by averaging the
tunneling amplitude subject to evolution due to the Hamiltonian
(\ref{Eq:M:H}) without the scattering term. This procedure is
similar to the one used in Ref.~\onlinecite{Keldysh} to study
ionization of quantum dot in high-frequency harmonic field. The
amplitude for an electron to tunnel from impurity site to the 2DEG
states is $\sum\limits_k \Delta \langle \bar d(t)\bar c_k^\dag
(t)\rangle$. Here, the overbar denotes evolution due to
Hamiltonian (\ref{Eq:M:H}) with only the scattering term absent.
The first nonvanishing order (with the complex prefactor ignored)
is
\begin{equation}\label{Eq:AppAv:TP2}
\int\limits_{ - \infty }^\infty  {dt'} \sum\limits_k {\Delta ^2 }
\langle T \tilde{d}(t)\tilde{c}_k^\dag  (t)\tilde{d}^\dag
(t')\tilde{c}_k (t')\rangle,
\end{equation}
where $\tilde{d}(t) = d \, e^{ - iE_0 t - i\frac{{V_{00}
}}{{\omega _0 }}\sin \omega _0 t}$ and $\tilde{c}_k (t) = c_k e^{
- iE_k t - i\frac{{V_{2DEG} }}{{\omega _0 }}\sin \omega _0 t}$.
These expressions are substituted into Eq.~(\ref{Eq:AppAv:TP2}) to
obtain
\begin{eqnarray}\label{Eq:AppAv:TP3}
\int\limits_{ - \infty }^\infty  {dt'} \sum\limits_k {\Delta ^2 }
\langle T d(t) c_k^\dag  (t) d^\dag  (t') c_k (t')\rangle
\\
\times \sum\limits_{nn'} {J_n (\lambda )J_{n'} (\lambda )} e^{ -
in\omega _0 t} e^{in'\omega _0 t'},
\end{eqnarray}
expanding the oscillatory exponents into the Bessel series. Here,
$d(t) = d \, e^{ - iE_0 t}$ and $c_k (t) = c_k e^{ - iE_k t}$. The
amplitude is then averaged over the fast driving field
oscillations yielding
\begin{equation}\label{Eq:AppAv:TP4}
\sum\limits_n {J_n^2 (\lambda )} \!\!\! \int\limits_{ - \infty
}^\infty \!\!{dt'} \sum\limits_k {\Delta ^2 } \langle T d(t)
c_k^\dag (t) d^\dag (t') c_k (t')\rangle e^{in\omega _0 \left( {t'
- t} \right)}.
\end{equation}
After the integration, one finally obtains the rates proportional
to
\begin{equation}\label{Eq:AppAv:TP5}
\mathrm{Im}\sum\limits_n {\Delta^2 J_n^2 (\lambda )} \sum\limits_k
{\frac{\theta(\pm 1) \mp n_k}{{E_0  - E_k  - \omega _0 n \pm
i0}}},
\end{equation}
which gives Eq.~(\ref{Eq:NST:SelfE}) with $\omega \to E_0$.


\end{document}